\documentclass[11pt,a4paper]{article}
\usepackage{jheppub}

\usepackage{amsmath}
\usepackage{amssymb}
\usepackage{graphicx}
\usepackage{bbm}
\usepackage{color}

\usepackage{indentfirst}
\usepackage{amsfonts}
\usepackage{caption}
\usepackage{subcaption}
\usepackage{hyperref}

\usepackage{slashed}

\usepackage{tikz}
\usetikzlibrary{arrows,positioning,shapes,mindmap}
\usetikzlibrary{decorations.pathmorphing}
\usetikzlibrary{decorations.markings}
\def\lsim{\mathrel{\raise.3ex\hbox{$<$\kern-.75em\lower1ex\hbox{$\sim$}}}}
\def\gsim{\mathrel{\raise.3ex\hbox{$>$\kern-.75em\lower1ex\hbox{$\sim$}}}}

\def\beq{\begin{equation}}
	\def\eeq{\end{equation}}
\def\bea{\begin{eqnarray}}
	\def\eea{\end{eqnarray}}

\def\bit{\begin{itemize}}
	\def\eit{\end{itemize}}
\def\bec{\begin{center}}
	\def\eec{\end{center}}

\title{Resonant Higgs pair production as a probe of stop at the LHC}% Force line breaks with \\
\author{ Guang Hua Duan$^{3,4}$}
\author{ Lei Wu$^{1,2}$}
\author{ Rui Zheng$^{5}$}
\affiliation{
	$^1$ Department of Physics and Institute of Theoretical Physics, Nanjing Normal University, Nanjing, Jiangsu 210023, China\\
	$^2$ ARC Centre of Excellence for Particle Physics at the Terascale, School of Physics, The University of Sydney, NSW 2006, Australia\\
	$^3$ Institute of Theoretical Physics, Chinese Academy of Sciences, Beijing 100190, China\\
	$^4$ School of Physical Sciences, University of Chinese Academy of Sciences, Beingjing 100049, China\\
	$^5$ Department of Physics, University of California, Davis, CA 95616, USA
}%
\emailAdd{ghduan@itp.ac.cn}
\emailAdd{leiwu@itp.ac.cn}
\emailAdd{ruizh@ucdavis.edu}

\abstract{Searching for top squark (stop) is a crucial task of the LHC. When the flavor conserving two body decays of the stop are kinematically forbidden, the stops produced near the threshold will live long enough to form bound states which subsequently decay through annihilation into the Standard Model (SM) final states. In the region of stop mixing angle $\theta_{\tilde{t}} \to 0$ or $\pi/2$, we note that the LHC-13 TeV diphoton resonance data can give a strong bound on the spin-$0$ stoponium ($\eta_{\tilde{t}}$) and exclude the constituent stop mass $m_{\tilde{t}}$ up to about 290 GeV. While in the large stop mixing region, the stoponium will dominantly decay to the Higgs pair. By analyzing the process $pp \to \eta_{\tilde{t}} \to h (\to b\bar{b} )h (\to \tau^+\tau^-)$, we find that a large portion of the parameter space on the $m_{\tilde{t}_1}$ - $\theta_{\tilde{t}}$ plane can be probed at $2\sigma$ significance level at the LHC with the luminosity ${\cal L}=3000$ fb$^{-1}$.}

\begin{document}

\maketitle

%%%%%%%%%%%%%%%%%%%%%%%%%%%%%%%%%%%%%%%%
\section{Introduction}
%%%%%%%%%%%%%%%%%%%%%%%%%%%%%%%%%%%%%%%%

Since the discovery of the Higgs boson at the Run 1 of the Large Hadron Collider (LHC) in 2012 \cite{higgs-atlas,higgs-cms}, the persuit of physics beyond the SM (BSM) becomes the primary goal in particle physics community. One of the most important guidelines in this endeavor is the famous naturalness principle which states that the physics at weak scale should be insensitive to quantum effects from much higher scales. Among all the proposed scenarios, supersymmetry (SUSY) remains as one of the most popular models, in which the quantum correction to the Higgs mass from the top quark is canceled by that from the stop. In this regard, the search for stop \cite{nsusy-1,nsusy-2,stop-2,stop-4,stop-6,stop-7,stop-8,stop-9,stop-12,stop-13,stop-14,stop-15,Batell:2015koa,Han:2013sga} is an important direction of testing SUSY naturalness at the LHC . %The discovery of stop is very likely to be the first signal of supersymmetry.

Till now, numerous efforts have been dedicated to the searching for stop in the LHC experiments. The experimental signatures of stop pair production depend on the stop-LSP mass splitting which leads to different decay modes. For instance, when $ m_{\tilde{t}_{1}} > m_t + m_{\tilde{\chi}^0_1}$ and $\tilde{t}_1$ mainly decays to $t \tilde{\chi}^0_1$, the top quark from stop decay can be quite energetic and a stop mass up to 940 GeV for a massless lightest neutralino has been excluded by the very recent LHC run-2 data \cite{ATLAS:2017msx}. When the flavor-conserving two body decays channels like $\tilde{t}_1 \to t \tilde{\chi}^0_1$ and $\tilde{t}_1 \to b \tilde{\chi}^+_1$ are kinematically forbidden, the primary decay channels of the light stop would be the three-body decay $\tilde{t}_1\to W^+ b \tilde{\chi}^0_1$, the two-body flavor-changing decay $\tilde{t}_1 \to c \tilde{\chi}^0_1$ or the four-body decay $\tilde{t}_1 \to bf'\bar{f}\tilde{\chi}^0_1$  \cite{hikasa,djouadi,margrate,andreas}. The current null results of LHC searches for these decay channels have correspondingly excluded the stop mass up to $\sim 500$~GeV, $310$~GeV and $370$~GeV for certain mass splitting between the stop and the LSP ~\cite{ATLAS:2017msx}.

It should be mentioned that such a light stop usually has very small decay width \cite{Martin:2008sv} compared to the typical binding energy of $\tilde{t}_1\tilde{t}_1^*$ bound state (stoponium). In this case, two stops produced near-threshold could live long enough to form a stoponium due to the Coulomb-like attraction via the QCD interaction. In contrast to the existing direct stop pair searches, stoponium if formed, will resonantly decay to a pair of the SM particles and can be independent of the assumptions of the LSP mass and the branching ratios of the stop. Therefore, it is expected that the search of stoponium can provide a complementary probe to the direct stop pair production at the LHC.

The phenomenologies of the stoponium have been studied at colliders \cite{Martin:2008sv,Drees:1993uw,Drees:1993yr,Bodwin:2016whr,Luo:2015yio,Ito:2016qsm,Barger:2011jt,Kim:2014yaa,Kang:2016wqi}. In particular, the diphoton channel was studied and found to be a promising way to observe stoponium at the LHC in Refs.~\cite{Martin:2008sv,Drees:1993uw,Drees:1993yr}. The diboson decay of stoponium with $WW$ and $ZZ$ final states were also examined in \cite{Barger:2011jt,Kim:2014yaa}.
In \cite{Kumar:2014bca}, the authors investigated the di-Higgs decay of stoponium with $b\bar{b}\gamma\gamma$ final states and found it to be a viable channel at the LHC. But the loop induced diphoton decay of the Higgs boson can be sizably affected by other sparticles, such as the light stau in the MSSM~\cite{carena}.

In this paper, we first confront the stoponium with the recent data of searching for high mass resonances at 13 TeV LHC. Then we explore the potential of probing the stop in Higgs pair production with $b\bar{b}\tau^+\tau^-$ final states at high-luminosity LHC (HL-LHC). As a comparison with $b\bar{b}\gamma\gamma$ channel, although the $b\bar{b}\tau^+\tau^-$ channel suffers from relatively complicated backgrounds, it has a larger branching ratio. Besides, it is expected that the reconstruction efficiency of $\tau$ can reach $\sim 80\%$ with the likelihood $\tau$ taggers in the future LHC experiment \cite{dolan,atlas}. This will make $b\bar{b}\tau^+\tau^-$ channel become another promising way of discovering, or confirming the stoponium at the LHC. The paper is organized as follows. In Sec.~\ref{sec:limit}, we introduce productions and decays of the stoponium and display the limits on stoponium mass from the LHC-13 TeV data. In Sec.~\ref{sec:simulation}, we investigate the observability of the di-Higgs decay of the stoponium with $b\bar{b}\tau^-\tau^+$ final states at the LHC. Finally, we draw our conclusions in Sec.~\ref{sec:conclusion}.

%%%%%%%%%%%%%%%%%%%%%%%%%%%%%%%%%%%%%%%%
\section{Diphoton resonance constraint on the stoponium}\label{sec:limit}
%%%%%%%%%%%%%%%%%%%%%%%%%%%%%%%%%%%%%%%%
In the gauge-eigenstate basis, the stop mass matrix is given by
\begin{eqnarray}
	M_{\tilde{t}}^2=
	\left(
	\begin{array}{cc}
		m_{\tilde{t}_L}^2 &m_tX_t^{\dag}\\
		m_tX_t& m_{\tilde{t}_R}^2\\
	\end{array}
	\right)
	\label{masssq}
\end{eqnarray}
with
\begin{eqnarray}
	&&m_{\tilde{t}_L}^2=m_{\tilde{Q}_{3L}}^2+m_t^2+m_Z^2\left(\frac{1}{2}-\frac{2}{3}\sin^2\theta_W\right)\cos2\beta,\\
	&& m_{\tilde{t}_R}^2=m_{\tilde{U}_{3R}}^2+m_t^2+\frac{2}{3}m_Z^2\sin^2\theta_W\cos2\beta,\\
	&& X_t = A_t -\mu \cot\beta,
\end{eqnarray}
where $m_{\tilde{Q}_{3L}}$ and $m_{\tilde{U}_{3R}}$ denote the soft-breaking mass parameters of the third generation left-handed squark doublet $\tilde{Q}_{3L}$ and the right-handed stop $\tilde{U}_{3R}$, respectively. $A_t$ is the soft-breaking trilinear parameter. We neglect the generation mixing in our study. The hermitian matrix Eq.~(\ref{masssq}) can be diagonalized by
a unitary transformation:
\begin{eqnarray}
	\left(
	\begin{array}{c}
		\tilde{t}_1 \\
		\tilde{t}_2 \\
	\end{array}
	\right)
	=
	\left(
	\begin{array}{cc}
		\cos\theta_{\tilde{t}} &\sin\theta_{\tilde{t}}\\
		-\sin\theta_{\tilde{t}}& \cos\theta_{\tilde{t}}\\
	\end{array}
	\right)
	\left(
	\begin{array}{c}
		\tilde{t}_L \\
		\tilde{t}_R \\
	\end{array}
	\right),
\end{eqnarray}
where $\theta_{\tilde{t}}\in [0,\pi)$ is the mixing angle between left-handed ($\tilde{t}_L$) and right-handed ($\tilde{t}_R$) stops. A very narrow decay width of stop \footnote{If the stop has a large decay width, it could in general produce a wide resonance signal and will be hardly observed on top of the continuum background \cite{Kats:2009bv}.} can naturally appear in the compressed region, in which the decay width of stop is suppressed either by phase space or loop factor. If the $\Gamma_{\tilde{t}_1}$ is much smaller than binding energy, stop pair produced near the threshold could form a bound state due to the strong attractive force mediated by gluons. Then, these bound states will proceed annihilation decay rather than the prompt decay of the constituent stop.

The production of stoponium is mainly from the gluon fusion at the LHC. In narrow-width approximation, the leading order (LO) cross section of stoponium is be given by~\cite{Martin:2008sv}
\begin{equation}
	\sigma( gg \to \eta_{\tilde{t}})=\frac{\pi^2}{8 m_{\eta_{\tilde{t}}}^3}\Gamma_{\eta_{\tilde{t}}\to g g}\frac{\hat{s}}{s}\int_{\frac{\hat{s}}{s}}^{1}\frac{d\,x}{x}f_g(x)f_g(\frac{\hat{s}}{x s})
\end{equation}
where $\hat{s}$ is squared center-of-mass energy at the parton level and is taken as $\hat{s}=m_{\eta_{\tilde{t}}}^2$ in our calculation. $\Gamma_{\tilde{\eta}_{\tilde{t}} \to gg}$ is the width of stoponium decay to di-gluon. The next-to-leading order QCD radiative corrections to stoponium production have been calculated in \cite{Hagiwara}. We include these effects by using the values of $K$-factor given in \cite{Younkin:2009zn}.

It should be noted that there are two main uncertainties in the computation of stoponium production rate. One of them lies in the parametrization of the wavefunction, which depends on the choice of QCD scale parameter $\Lambda$ \cite{Hagiwara:1990sq}. Larger value of $\Lambda$ leads to greater coupling and hence stronger binding between the constituent stops. We adopt $\Lambda=300$~MeV by following~\cite{Younkin:2009zn}. The other uncertainty comes from the contributions of excited bound states, such as $nS(n\geq2)$ and $1P$ states. In particular, the effects of higher $S$-wave states are compared in \cite{Younkin:2009zn}. The excited states can contribute by either first decaying into the lowest stoponium state ($1S$) or decaying directly into SM final states. For instance, the non-annihilation decay of the $2S$ state could go entirely to the $1S$ state and the signal could be merged with that of the ground state due to the detector energy resolution \cite{Martin:2008sv}. In general, states with different angular momentum could have very distinct decay modes. Without thorough knowledge of the decay modes, we will take a conservative approach and focus on the $1S$ state.

The main decay channels of the stoponium include $\eta_{\tilde{t}} \to \gamma\gamma, \gamma Z, ZZ, WW, gg, hh, t\bar{t}$.  The LO partial decay widths into transverse gauge bosons are \cite{Martin:2008sv}
\begin{equation}\label{eq:widthgg}
	\Gamma(\eta_{\tilde{t}}\to gg) \simeq \frac{4}{3}\alpha_S^2\frac{|R(0)|^2}{m_{\eta_{\tilde{t}}}^2},\qquad \Gamma(\eta_{\tilde{t}}\to \gamma\gamma) \simeq \frac{32}{27}\alpha^2\frac{|R(0)|^2}{m_{\eta_{\tilde{t}}}^2}
\end{equation}
where $R(0)=\sqrt{4\pi}\psi(0)$ is the radial wavefunction at the origin. In the nonrelativistic limit ($v\to 0$), only four-point interaction contributes to the stoponium decays $\eta_{\tilde{t}} \to gg,\gamma\gamma$. All other decay widths can be found in \cite{Drees:1993yr,Kim:2014yaa}. Radiative corrections to stoponium annihilation decays to hadrons, photons, and Higgs bosons were calculated in Ref.~\cite{Martin:2009dj}.

\begin{figure}
	\centering
	\includegraphics[scale=0.5]{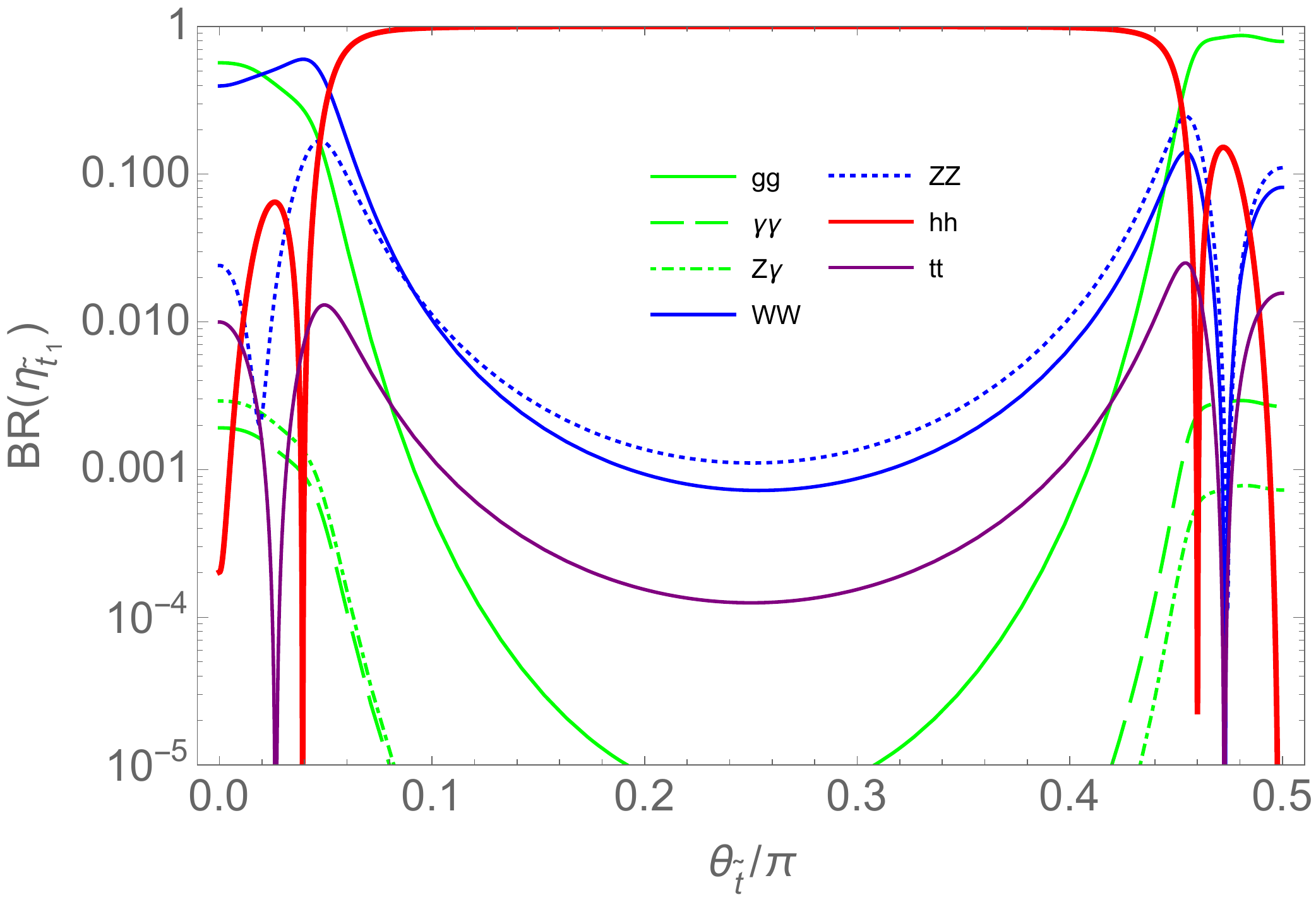}
	\vspace{-0.5cm}
	\caption{The decay branching ratios of the stoponium with respect to the mixing angle $\theta_{\tilde{t}}$. Here we take $\tan\beta=10$, $m_{\tilde{t}_1}=0.2$~TeV and $m_{\tilde{t}_2}=2$~TeV for example. Note that the branching ratios are symmetric about $\theta_{\tilde{t}}=\pi/2$, we plot only the region $\theta_{\tilde{t}}\in [0,\pi/2]$ here and also in Fig.~\ref{fig:contour}. }
	\label{fig:br}
\end{figure}
In Fig.~\ref{fig:br}, we display the decay branching ratios of the stoponium with respect to the mixing angle $\theta_{\tilde{t}}$, where we assume $\tan\beta=10$, $m_{\tilde{t}_1}=0.2$ TeV and $m_{\tilde{t}_2}=2$ TeV. It can be seen that the stoponium dominantly decays to di-gluon when the mixing angle $\theta_{\tilde{t}}$ approaches $0$ or $\pi/2$. While if $\tilde{t}_L$ and $\tilde{t}_R$ have a sizable mixing, the stoponium will dominantly decay to a pair of Higgs bosons because of the enhancement induced by the Higgs-stop coupling $\lambda_{h\tilde{t}_1\tilde{t}_1}$ \footnote{The trilinear coupling between the SM Higgs and stop quark $\tilde{t}_1$ takes the form \cite{Batell:2015zla}: $\lambda_{h\tilde{t}_1\tilde{t}_1}=\sqrt{2}v\Bigg(\frac{m_t^2}{v^2}+\frac{m_Z^2c_{2\beta}}{v^2}\Big[c_t^2(\frac{1}{2}-\frac{2}{3}s_W^2)+s_t^2(\frac{2}{3}s_W^2)\Big]+s_t^2c_t^2\frac{m_{\tilde{t}_1}^2-m_{\tilde{t}_2}^2}{v^2}\bigg)$.}. We also checked and found that branching ratios of the stoponium have a weak dependence of $\tan\beta$. So we will assume $\tan\beta=10$ in our following calculations. Due to the distinctive signature of two photon final states, the stoponium decay to diphoton offers a very sensitive way to observing stoponium at hadron colliders.

%\begin{equation}
%\lambda_{h\tilde{t}_1\tilde{t}_1}=\sqrt{2}v\Bigg(\frac{m_t^2}{v^2}+\frac{m_Z^2c_{2\beta}}{v^2}\Big[c_t^2(\frac{1}{2}-\frac{2}{3}s_W^2)+s_t^2(\frac{2}{3}s_W^2)\Big]+s_t^2c_t^2\frac{m_{\tilde{t}_1}^2-m_{\tilde{t}_2}^2}{v^2}\bigg)
%\end{equation}

%\begin{equation}
%E_b=\frac{C^2\bar{\alpha}_S^2 m_{\tilde{t}}}{4},\qquad a_0=\frac{2}{C \bar{\alpha}_S m_{\tilde{t}}},\qquad |\psi(0)|^2=\frac{1}{\pi a_0^3}=\frac{(C \bar{\alpha}_S m_{\tilde{t}})^3}{8\pi}
%\end{equation}
%The potential energy in the case of stoponium is not coulomb like. The wavefunction will be different, see Refs[cite].

%Wavefunction Corrections:\\
%Refer to NPB344(1990)1 [cite], Appendix A.
%
%
%For $(3\times \bar{3})_1$ quarkonia,
%\begin{align}
%&M(nS)=2 m_Q+a_0+a_1 x+a_3 x^3\\
%&\frac{|R_{nS}(0)|^2}{M(nS)^2}=b_0+b_1x+b_2 x^2+b_3 x^3
%\end{align}
%with $x=\log[m_Q(\text{GeV})/25]$. The values of the constants can be found in the appendix of [cite].
%
%
%
%QCD Corrections:\\
%Refer to 0901.4318 and 0912.4813
%
%%%%%%%%%%%%%%%%%%%%%%%%%%%%%%%%%%%%%%%%
%\section{Results}
%%%%%%%%%%%%%%%%%%%%%%%%%%%%%%%%%%%%%%%%
%\section{Current Limit on Stoponium}\label{sec:limit}

\begin{figure}
	\centering
	\includegraphics[scale=0.5]{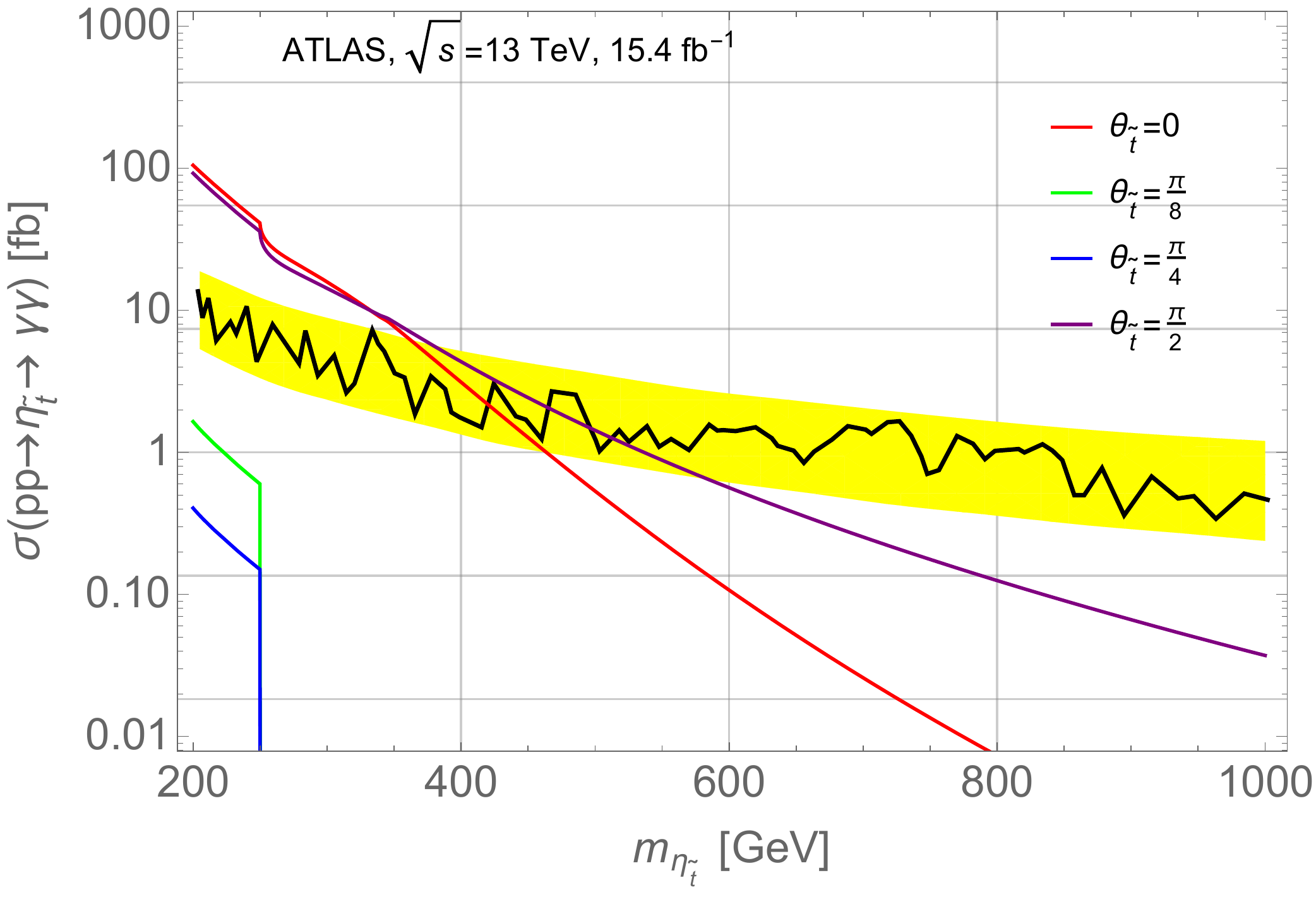}
	\vspace{-0.5cm}
	\caption{Constraint on the stoponium from the LHC-13 TeV diphoton resonance data. The $2\sigma$ experimental upper limit (yellow band) is taken from \cite{diphotonconstraint}. Here we also assumed $\tan\beta=10$ and $m_{\tilde{t}_2}=2$~TeV.}
	\label{fig:constraint}
\end{figure}

The bound on stoponium from 8~TeV run at the LHC is given in \cite{Batell:2015zla}. In Fig.~\ref{fig:constraint}, we update the result with the LHC-13 TeV diphoton resonance data~\cite{diphotonconstraint}. We can see that the stoponium mass can be excluded up to about 580 GeV for the mixing angles $\theta_{\tilde{t}}=\pi/2$, which is stronger than that from LHC-13 TeV direct searches for the four-body decay $\tilde{t}_1 \to bf'\bar{f}\tilde{\chi}^0_1$ with pure bino LSP in the region of $m_{\tilde{t}_1}-m_{\tilde{\chi}^0_1}<15$ GeV~\cite{ATLAS:2017msx}. However, due to the branching ratio suppression effect, there is still no constraint on the stoponium from the diphoton data for the mixing angles $\theta_{\tilde{t}}=\pi/8,\pi/4$. We also checked the bounds on the stoponium from current null results of LHC searches for $Z\gamma$ and diboson resonances and found that they can not give stronger limits than the diphoton data.

\section{Di-Higgs decay of stoponium with $b\bar{b}\tau^+\tau^-$ final states at the LHC}\label{sec:simulation}
%%%%%%%%%%%%%%%%%%%%%%%%%%%%%%%%%%%%%%%%

Given that the stoponium can have a large branching fraction into the two Higgs bosons, we will investigate its observability through the resonant Higgs pair production with $b\bar{b}\tau^+\tau^-$ final states at the 14~TeV LHC,
\begin{equation}
	pp \to \eta_{\tilde{t}} \to hh \to b\bar{b}\tau^+\tau^-.
\end{equation}
where one tau lepton decays hadronically ($\tau_{had}$) and the other decays leptonically. $\tau_{had}$ is reconstructed using clusters in the electromagnetic and hadronic calorimeters with medium criterion \cite{Aad:2014rga}.

We generate parton-level events of the stoponium production and subsequent decay into Higgs pair using the code for resonant Higgs pair production \cite{Frederix:2014hta} within \textsf{MG5$\_$aMC@NLO} \cite{Alwall:2014hca}, in which $\tau$ lepton decays are modeled by \textsf{TAUOLA} \cite{Jadach:1993hs}. Then we perform parton shower and hadronization with \textsf{PYTHIA} \cite{pythia}. The fast detector simulation is implemented with \textsf{Delphes}~\cite{delphes}. We use the $b$-jet tagging efficiency parametrization as 80$\%$ \cite{cms-b} and set the misidentification $10\%$ and $1\%$ for $c$-jets and light jets, respectively. We also assume the $\tau$ tagging efficiency is $40\%$. We set the renormalization scale $\mu_R$ and factorization scale $\mu_F$ as the default event-by-event value. We cluster the jets by choosing the anti-$k_t$ algorithm with a cone radius $\Delta R=0.4$ \cite{anti-kt}. The major backgrounds come from events with a jet misidentified as $\tau_{had}$, including $t\bar{t}$, $Z(\to \tau^+\tau^-)b\bar{b}$ and $Z(\to \tau^+\tau^-)jj$ processes.

\begin{figure}[t]
	\centering
	\includegraphics[width=6in]{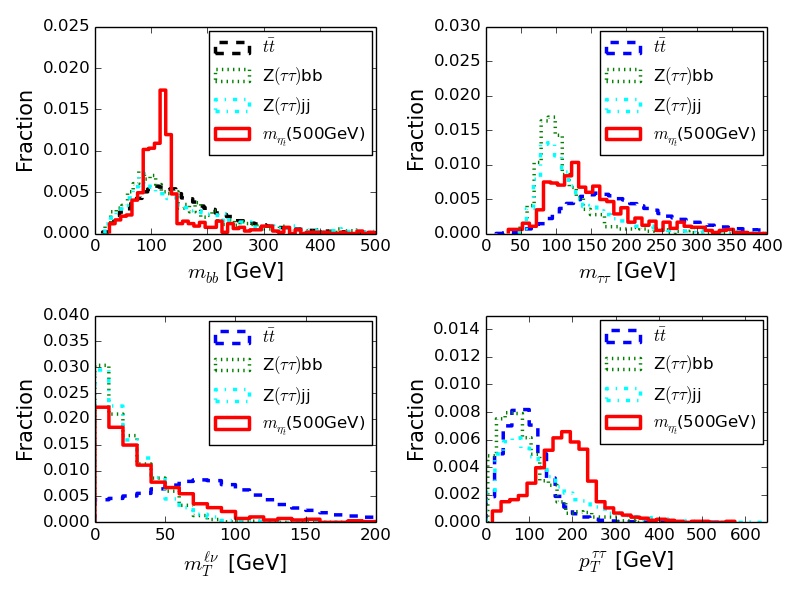}
	\vspace{-0.5cm}
	\caption{Distributions of the di-tau invariant mass $m_{\tau\tau}$, two $b$-jets invariant mass $m_{bb}$, the transverse mass of the lepton plus missing energy system $m_T^{\ell \nu}$ and the di-tau transverse momentum $p^{\tau\tau}_T$. The stoponium mass is taken as $m_{\eta_{\tilde{t}}}=500$ GeV.}
	\label{fig:distribution}
\end{figure}

%
%$80<m_{bb}<150$ GeV, $80<m_{\tau\tau}<150$ GeV, $m^{\ell \nu}_T<50$ GeV, and $p_{T}^{\tau\tau}>120$GeV.
In Fig.~\ref{fig:distribution}, we present distributions of the di-tau invariant mass $m_{\tau\tau}$, two $b$-jets invariant mass $m_{bb}$, the transverse mass of the lepton plus missing energy system $m_T^{\ell \nu}$ and the di-tau transverse momentum $p^{\tau\tau}_T$. The simple transverse mass method is used to reconstruct $m_{\tau\tau}$ from the observed lepton, $\tau_{had}$ and $E^{miss}_T$. One can see that $m_{\tau\tau}$ distribution shows a relatively broad peak around the Higgs boson mass with a long tail \footnote{This can be improved by using the advanced experimental MMC reconstruction technique \cite{ditau}.}, as a comparison with $m_{b\bar{b}}$ distribution. Another variable $m^{\ell\nu}_{T}$ can effectively reduce $t\bar{t}$ background since the lepton in signal is not from $W$ boson decay. The variable $p^{\tau\tau}_T$ is used to select the events with the boosted Higgs boson candidate on the transverse plane. For such events, $m_{\tau\tau}$ resolution is improved and a better separation between the signal $\eta_{\tilde{t}} \to \tau\tau$ and the background $Z \to \tau\tau$ is achieved. This selection also has the advantage
of reducing the QCD multijet background.

%\begin{figure}[h]
%  	\centering
%  	\includegraphics[scale=0.4]{mbbll.jpg}
%  	\vspace{-0.5cm}
%  	\caption{Distributions of $m_{bb\tau\tau}$. The signal benchmark point are
%  		$m_{\eta_{\tilde{t}}}=350,500,800$ GeV.}
%  	\label{fig:mbbll}
%\end{figure}

In our analysis, we select events that satisfy the following criteria:
\begin{itemize}
	\item We require exactly one lepton ($e$ or $\mu$) with $p_{T} (\ell) >26$ GeV , $|\eta_e|< 2.47$ or $|\eta_\mu|< 2.5$. We further require the presence of a hadronically decayed tau $\tau_h$ carrying opposite electric charge with $p_{T}(\tau_h)>20$~GeV and $|\eta_{\tau_h}|< 2.5$.
	\item  We require at least two jets with $p_T(j)>30$~GeV  and $|\eta_{j}|< 2.5$ and two of them are b tagged.
	\item  We require 80 GeV $<m_{bb}<$ 150 GeV, 80 GeV $<m_{\tau\tau}<$ 150 GeV, $m^{\ell\nu}_T<50$ GeV, $p_{T}^{\tau\tau}>120$ GeV and $|m_{bb\tau\tau}-m_{\eta_{\tilde{t}}}|<0.08 m_{\eta_{\tilde{t}}}$.
	%  	
	%  	%$|m_{bb\tau\tau}-m_{\eta_{\tilde{t}}}|<40$GeV for $m_{\eta_{\tilde{t}}}<600$ GeV and $m_{bb\tau\tau}>650$GeV for $m_{\eta_{\tilde{t}}}>600$ GeV.
\end{itemize}

\begin{table}[ht!]
	\caption{Cut flow analysis of the cross sections (fb) for the signal and backgrounds at 14 TeV LHC. The benchmark point is chosen as $m_{\eta_{\tilde{t}}}=500$ GeV and $\sigma(gg \to \eta_{\tilde{t}} \to hh) = 1$ pb.}
	\footnotesize\begin{tabular}{|c|c|c|c|c|c|c|}
		\hline
		Cuts &  $m_{bb}$& $m_{\tau\tau}$ & $m^{\ell \nu}_T$ & $p_{T}^{\tau\tau}$ &$|m_{bb\tau\tau}-m_{\eta_{\tilde{t}}}|$  \\
		&$\in[80,150]$ GeV&$\in[80,150]$ GeV&$<50$GeV&$>120$GeV&$<0.08m_{\eta_{\tilde{t}}}$\\
		\hline
		$t\bar{t}$ &445.48& 128.79 & 55.32&12.46&0.29 \\
		\hline
		
		$Z(\tau\tau)bb$ &7.40&5.35&4.70&0.62& $<0.02$\\
		\hline
		$Z(\tau\tau) jj$&11.87&7.92&7.04&1.62& 0.13\\
		\hline
		signal($m_{\eta_{\tilde{t}}}=500$ GeV) & 1.55& 0.82& 0.64&0.54&0.25 \\
		\hline
	\end{tabular}
	\label{tab:cutflow}
\end{table}

In Table~\ref{tab:cutflow}, we present a cut flow of cross sections for the signal and backgrounds at 14 TeV LHC. After the di-$b$ jets and di-tau invariant mass cuts, we find that the cut $m^{\ell\nu}_{T}<50$ GeV can reduce the $t\bar{t}$ background by about half. The cut $p^{\tau\tau}_T>120$ GeV can suppress $Z(\to \tau\tau)jj$ and $Z(\to \tau\tau)bb$ backgrounds by an extra factor of six. The total invariant mass cut $|m_{bb\tau\tau}-m_{\eta_{\tilde{t}}}|<0.08m_{\eta_{\tilde{t}}}$ can further hurt $t\bar{t}$ background by about ${\cal O}(10^2)$ and $Z(\to \tau\tau)jj$ and $Z(\to \tau\tau)bb$ by about ${\cal O}(10)$.

\begin{figure}[h]
	\centering
	\includegraphics[width=4in]{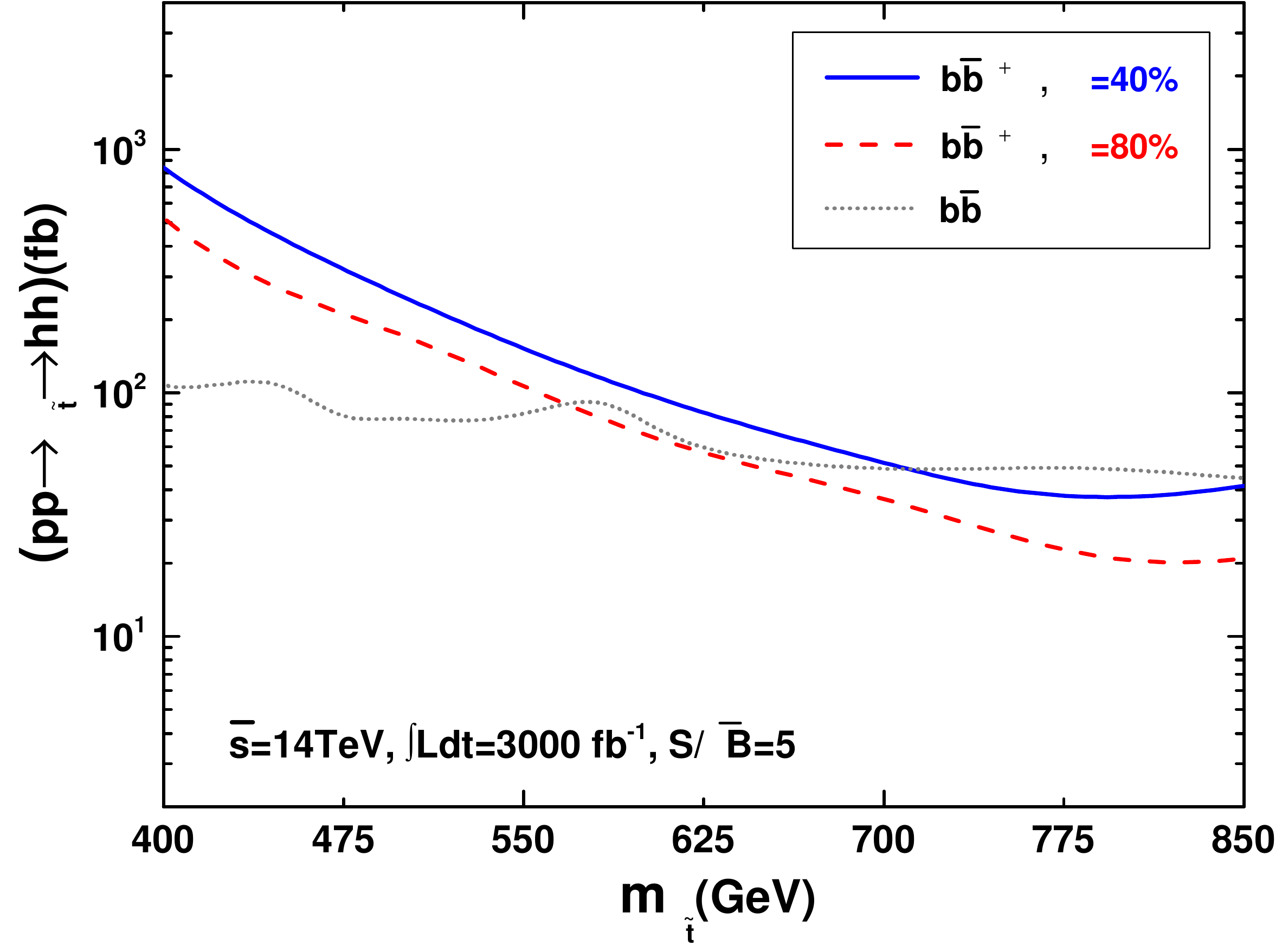}
	\caption{Cross sections of the process $pp \to \eta_{\tilde{t}} \to hh$ with $b\bar{b}\tau^+\tau^-/b\bar{b}\gamma\gamma$ final states needed for the signal significance $S/\sqrt{B}=5 \sigma$ at the HL-LHC. The result for $b\bar{b}\gamma\gamma$ final state is taken from Ref.~\cite{Kumar:2014bca}}
	\label{fig:limit}
\end{figure}

In Fig.~\ref{fig:limit}, we plot the cross sections of the process $pp \to \eta_{\tilde{t}} \to hh$ with $b\bar{b}\tau^+\tau^-/b\bar{b}\gamma\gamma$ final states needed for the signal significance $S/\sqrt{B}=5\sigma$ at the HL-LHC. It can be seen that the cross section of the process $pp \to \eta_{\tilde{t}} \to hh \to b\bar{b}\tau^+\tau^-/b\bar{b}\gamma\gamma$ should be about 800 fb/100 fb to reach $5\sigma$ significance at $m_{\eta_{\tilde{t}}}=400$~GeV. When the stoponium is heavier than about 700 GeV, the required cross section of $b\bar{b}\tau^+\tau^-$ channel for a tau tagging efficiency $\epsilon_\tau=40\%$ can be comparable with that of $b\bar{b}\gamma\gamma$ channel studied in \cite{Kumar:2014bca}. If $\tau$ tagging efficiency can be improved to $\sim 80\%$ estimated in~\cite{dolan,atlas}, the sensitivity of $b\bar{b}\tau^+\tau^-$ channel is expected to become better than that of $b\bar{b}\gamma\gamma$ channel for $m_{\eta_{\tilde{t}}} \gtrsim 570$ GeV.

\begin{figure}
	\centering
	\includegraphics[scale=0.6]{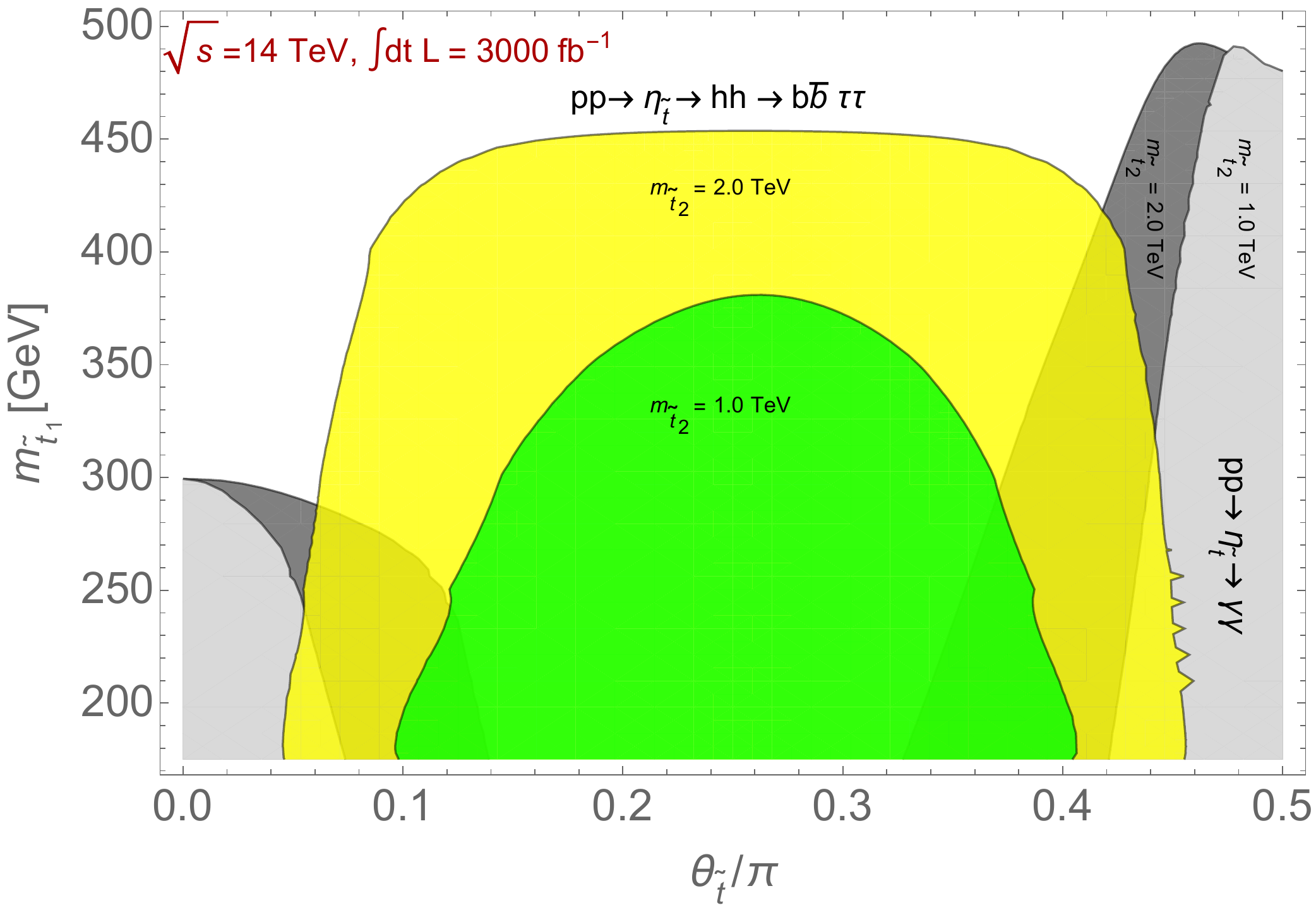}
	\caption{$2\sigma$ exclusion limits from the di-Higgs decay channel $\eta_{\tilde{t}} \to hh \to b\bar{b}\tau^+\tau^-$ and the di-photon decay channel $\eta_{\tilde{t}} \to \gamma\gamma$ for $m_{\tilde{t}_2}=1$ TeV and 2~TeV on the plane of $m_{\tilde{t}_1}$ versus stop mixing angle $\theta_{\tilde{t}}$ at the HL-LHC. The result of di-photon decay channel is taken from Ref.~\cite{Batell:2015zla}.}
	\label{fig:contour}
\end{figure}

In Fig.~\ref{fig:contour}, we show $2\sigma$ exclusion limits from the di-Higgs decay channel $\eta_{\tilde{t}} \to hh \to b\bar{b}\tau^+\tau^-$ and the di-photon decay channel $\eta_{\tilde{t}} \to \gamma\gamma$ for $m_{\tilde{t}_2}=1$ TeV and 2~TeV on the plane of $m_{\tilde{t}_1}$ versus stop mixing angle $\theta_{\tilde{t}}$ at the HL-LHC. We can see that the stop mass $m_{\tilde{t}_1}$ can be excluded up to $\sim 380 (450)$~GeV in the large stop mixing region $\pi/7\lesssim \theta_{\tilde{t}} \lesssim \pi/3$ by the di-Higgs decay channel $\eta_{\tilde{t}} \to hh \to b\bar{b}\tau^+\tau^-$, since the branching ratio of $\eta_{\tilde{t}}\to hh$ depends on the Higgs-stop coupling $\lambda_{h\tilde{t}_1\tilde{t}^*_1}$. For a given mixing angle $\theta_{\tilde{t}}$, a larger $m_{\tilde{t}_2}$ sets a stronger bound on $m_{\tilde{t}_1}$ because the Higgs-stop coupling $\lambda_{h\tilde{t}_1\tilde{t}^*_1}$ is proportional to the mass difference $m^2_{\tilde{t}_1} - m^2_{\tilde{t}_2}$. The di-photon decay channel $\eta_{\tilde{t}} \to \gamma\gamma$ mainly excludes small stop mixing region, such as $\theta_{\tilde{t}} \lesssim \pi/7$ or $\theta_{\tilde{t}} \gtrsim \pi/3$, which is complementary to the di-Higgs decay channel.

%\newpage
%%%%%%%%%%%%%%%%%%%%%%%%%%%%%%%%%%%%%%%%

\section{Conclusions}\label{sec:conclusion}
%%%%%%%%%%%%%%%%%%%%%%%%%%%%%%%%%%%%%%%%
In this paper, we confront the stoponium with the recent data of searching for high mass resonances at 13 TeV LHC, and explore the potential of probing the stoponium in resonant Higgs pair production with $b\bar{b}\tau^+\tau^-$ final states at the LHC. We note that the LHC-13 TeV diphoton resonance data can give a strong bound on the spin-0 stoponium ($\eta_{\tilde{t}}$) and exclude the constituent stop mass $m_{\tilde{t}_1}$ up to about 290 GeV in the small stop mixing region. While in the large stop mixing region, the stoponium will dominantly decay to the Higgs pair. By analyzing the process $pp \to \eta_{\tilde{t}} \to h (\to b\bar{b} )h (\to \tau^+\tau^-)$, we find that the stop mass $m_{\tilde{t}_1}$ can be excluded up to $\sim 380~(450)$~GeV at the LHC with the luminosity ${\cal L}=3000$ fb$^{-1}$.

%%%%%%%%%%%%%%%%%%%%%%%%%%%%%%%%%%%%%%%%
\section*{Acknowledgements}
Lei Wu was supported by the National Natural Science Foundation of China (NNSFC) under grants No. 11705093, 11305049, and by the Australian Research Council.

%\section*{Appendix}
%-----------------------------------------------------------------------------------------------------------------------------
%\bibliographystyle{elsarticle-num}
%\bibliography{references}
%-----------------------------------------------------------------------------------------------------------------------------

%%%%%%%%%%%%%%%%%%%%%%%%%%%%%%%%%%%%%%%%%%%%%%%%%%%%%%
\end{document}